# TECHNICAL SCIENCES

## ANALYSIS OF AI EFFECTIVENESS IN REDUCING HUMAN ERRORS IN PROCESSING TRANSPORTATION REQUESTS

**Oleksandr Korostin**
*master's degree,*
*Classic Private University*
*Ukraine, 69002, Zaporozhye, Zhukovsky st., 70*
DOI: 10.5281/zenodo.13786097

**Abstract**
This article examines the characteristics of human errors in processing transportation requests. The role of artificial intelligence (AI) in maritime transportation is explored. The main methods and technologies used for automating and optimizing the handling of transportation requests are analyzed, along with their impact on reducing the number of errors. Examples of successful AI implementation in large companies are provided, confirming the positive influence of these technologies on overall operational efficiency and customer service levels.

**Introduction**

Maritime transport plays a crucial role in global trade, facilitating the movement of goods between continents and contributing to economic growth and international integration.

Maritime transport is a key channel for global supply chains, and its efficiency largely determines the stability and predictability of the economic sphere. Processing transportation requests in the maritime industry involves numerous complex stages, each requiring high accuracy and promptness, as even a minor error can lead to serious delays, additional costs, and a loss of customer trust. In a highly competitive environment and with the drive to optimize processes, companies must seek effective solutions capable of minimizing risks associated with the human factor.

One of the most effective directions in this area is the use of artificial intelligence (AI). Its implementation can significantly reduce the influence of human factors. Automating routine tasks, analyzing large volumes of data, and the ability of AI to predict and adapt to changes in real-time open new horizons for optimizing maritime transport processes. The aim of this article is to analyze the effectiveness of using AI in reducing human error in processing transportation requests.

**Main part. Human errors in maritime transportation**

According to statistics from the International maritime organization, water transport was the dominant mode of transportation for the USA in 2022. Vessels transported goods with a total weight of 1,59 billion short tons (fig. 1).

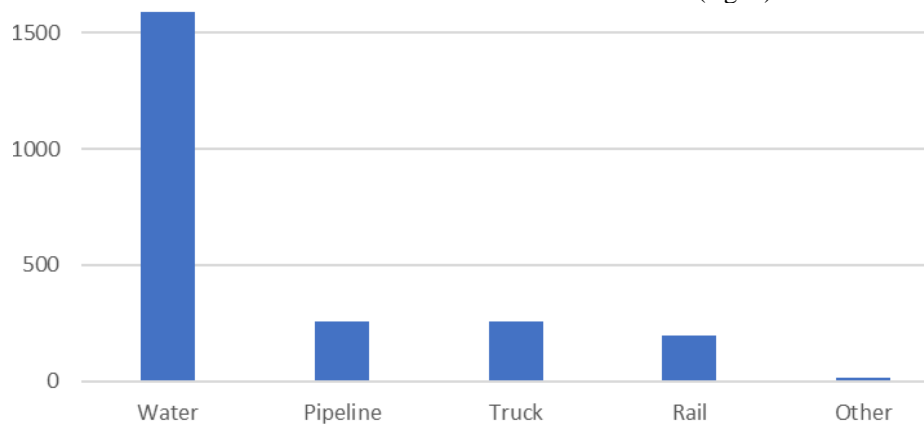

*Pic. 1. Weight of goods traded by the USA in 2022, broken down by mode of transport, in billion short tons [1]*

Maritime logistics constitutes a complex and multifaceted system, where each stage, from route planning to cargo delivery, demands a high degree of accuracy and coordination. In the context of intense work and constant pressure associated with international trade, human errors can lead to severe consequences. These errors not only affect the financial outcomes of companies but can also jeopardize the safety of the crew and the integrity of the cargo. Table 1 presents the main types of human errors, their causes, and potential consequences.

Table 1
**Types of human errors, causes of their occurrence, and potential consequences [2, 3]**

| Type of error | Causes of occurrence | Consequences |
|---|---|---|
| Documentation errors | Incorrect form filling, outdated information, lack of training. | Delays in cargo processing, fines, legal issues. |
| Cargo management errors | Improper cargo placement, errors in instructions, lack of oversight. | Damage or loss of cargo, increased transportation costs. |
| Navigation errors | Calculation errors in route planning, insufficient knowledge of navigation systems. | Delays, collision risks, loss of the vessel. |
| Coordination errors | Inadequate communication among process participants, incorrect data. | Delays, confusion in logistics, additional expenses. |
| Maintenance errors | Improper or insufficient maintenance of the vessel. | Breakdowns, accidents, disruption of schedules. |
| Personnel management errors | Lack of qualified personnel, insufficient training and workshops. | Decreased operational efficiency, increased number of incidents. |
| Compliance errors | Ignoring standards and regulatory requirements, lack of oversight. | Fines, legal sanctions, safety risks. |

In the author's opinion, effective management of maritime transportation requires consideration of numerous parameters related to the human factor. Major types of errors, such as documentation errors, cargo management errors, navigation errors, coordination errors, and maintenance errors, significantly impact the overall reliability and safety of operations. Understanding the causes of such issues is essential for identifying ways to minimize and prevent them, where the use of modern technologies plays a crucial role.

**The role of AI in maritime transportation.** In the context of technological progress, AI is an important tool transforming the approach to managing maritime operations. With its capabilities in big data analysis, process automation, and forecasting, AI is involved in various aspects of maritime logistics. Modern AI-based systems are capable of processing vast amounts of data received from various sensors and systems, enabling more effective forecasting and management of ship routes, analysis of weather conditions, and optimization of resource utilization. This not only reduces the risk of accidents and enhances safety but also helps minimize fuel and maintenance costs [4]. Through machine learning (ML) and analytics technologies, AI also improves cargo management by predicting needs and optimizing the placement of goods on board. As a result, maritime transportation becomes more flexible and adaptive to changing conditions and market demands. Thus, AI significantly contributes to the modernization of maritime logistics, facilitating safer, more efficient, and cost-effective operations, while opening new horizons for industry development.

**AI as a tool for reducing human errors in processing transport requests**

The implementation of AI systems helps reduce the likelihood of errors associated with the human factor by automating processes, improving data analysis, and enhancing accuracy. Its use offers a wide range of tools and technologies for minimizing human errors.

One such method is the **automation of request processing systems** in maritime transportation. These systems utilize AI and ML algorithms to process large volumes of logistics-related data, routes, and cargo. By automating processes such as route planning and documentation handling, the likelihood of human factor-related errors is reduced. This is achieved through automatic request routing, which eliminates the chance of incorrect task distribution [5]. Chatbots and virtual assistants that use natural language processing (NLP) algorithms can quickly and accurately handle standard requests, lowering the risk of mistakes in interpretation. The use of templates and automated responses ensures consistency and relevance of information, further helping to avoid inaccuracies.

Automation also allows for the integration of various data sources, including weather information and port congestion data. This enables the prediction of potential problems and proactive measures, minimizing errors in transport planning. As a result, companies can manage their resources more effectively, optimizing processes and reducing costs.

For **data verification and correction in maritime transportation**, AI is also becoming an essential tool for reducing human errors. In complex logistics environments, where numerous variables such as routes, cargo, and schedules are involved, AI aids in automating the information processing workflow. These systems can analyze large amounts of data and identify inconsistencies or errors that may occur during manual data entry [6]. For example, AI can automatically verify the accuracy of entered cargo data, such as weight, dimensions, and contents, comparing them against predefined standards. If discrepancies are detected, the system can not only signal an error but also suggest corrections, thereby preventing potential issues during transport.

The use of ML algorithms allows for the **analysis of historical data and identification of complex dependencies** that may be overlooked in traditional approaches. This enables accurate predictions of potential problems, such as delivery delays or disruptions in supply chains, and helps optimize planning processes [7]. These systems can adapt to changes in data and adjust forecasts in real time, providing more flexible and efficient transportation management.

The use of NLP makes it possible to **improve customer interaction** in the field of maritime transportation, which helps to reduce the number of human errors. In situations where customers submit diverse requests, NLP allows for the automatic processing and analysis of textual information, providing quick and accurate responses. This significantly reduces the likelihood of misunderstandings and errors that may arise from manual interpretation of messages.

Systems utilizing NLP technologies can recognize and interpret customer intentions, allowing them to respond effectively to queries. Furthermore, NLP can analyze customer feedback and comments to identify common issues or requests. This enables companies to enhance their services and adapt to customer needs, thereby reducing the likelihood of errors related to dissatisfaction or incorrect expectations [8]. The integration of NLP into customer interaction management systems leads to improved service quality and higher overall satisfaction.

The use of these AI tools significantly enhances the accuracy, timeliness, and efficiency of processes, reducing the risk of human errors and improving the overall quality of services and operations.

**Benefits of using AI to reduce human error in processing transport requests**

Human error correction tools using AI are a powerful resource that significantly improves the efficiency and safety of operations. In today's environment, where cargo volumes are increasing and service quality demands are becoming more stringent, the use of AI is not only practical but also essential. Table 2 presents the main advantages of this approach to correcting human errors, as well as the challenges that companies face when implementing AI in request processing.

Table 2

Advantages and limitations of using AI in request processing to reduce human error [9]

| | |
|---|---|
| **Advantages** | The ability to automate data processing processes and standardize work procedures, which reduces the likelihood of errors related to the human factor. |
| | The capacity to analyze large volumes of data and make predictions with high accuracy. |
| | Automation of routine tasks using AI accelerates data processing and reduces the time spent on task execution. |
| | Prompt handling of customer requests and offering personalized solutions, which improves service quality and customer satisfaction. |
| | The ability to detect anomalies and suspicious behavior, helping to prevent fraud and enhance quality control. |
| | Continuous improvement through learning from accumulated data and mistakes. |
| **Limitations** | The implementation of AI requires significant costs for development and installation. However, the long-term benefits often offset these initial investments. |
| | The need for maintenance and updates. |
| | The implementation of AI increases risks associated with cyber threats. However, modern cybersecurity measures and data protection methods can effectively address these risks, ensuring the security of the system. |
| | Integrating AI into existing processes can be complex. Ccareful planning and the use of experienced professionals can help overcome these challenges. |

In the author's opinion, the advantages provided by the use of AI tools for correcting human errors in maritime transportation significantly outweigh the disadvantages. It not only improves the efficiency and security of operations, but also helps companies adapt to rapidly changing market conditions. With the right approach to technology implementation and risk minimization, the use of AI becomes an important step towards modern and highly effective management of maritime transportation.

The predominance of positive aspects is confirmed by the real experiences of companies that have implemented AI to correct human errors. For instance, **Crowley Maritime** actively utilizes AI to optimize the processes of handling transport requests, which significantly reduces human errors. Through the use of AI, Crowley Maritime has been able to increase the accuracy of delivery time forecasting to 95%. The AI-based document management system has automated the processing of shipping invoices, leading to a 30% reduction in errors. In 2022, the time required for documentation processing decreased from 12 minutes to 4 minutes, significantly speeding up the entire transportation process [10].

Another company, the **Port of Los Angeles**, is actively implementing AI to improve the handling of transport requests, which in turn significantly reduces the number of human errors. For example, using AI, delay prediction systems were able to forecast the likelihood of delays with 90% accuracy. This allowed operators to take preemptive measures, reducing the chance of misunderstandings and errors in scheduling. In 2022, the port implemented automated document processing systems that use ML to extract and verify data, resulting in a 30% reduction in document-related errors [11].

Thus, through the implementation of AI, companies have significantly improved the efficiency of their operations, reduced the number of errors, and optimized the entire process of handling transport requests.

**Conclusions**

The introduction of AI in the field of maritime transportation represents a significant advancement in reducing human error in processing transport requests. With the help of AI, it is possible to analyze cargo data more accurately, predict needs and manage resources, which leads to a significant reduction in the number of errors during planning and processing. Platforms and tools utilizing ML and big data analytics ensure more transparent and efficient supply chain management, improving overall operational efficiency.

The application of AI in maritime transportation not only enhances customer service quality but also contributes to more efficient resource utilization, reducing costs and increasing profits. In an environment of growing competition and a dynamically changing market, companies that actively implement AI technologies will become industry leaders, promoting safer, more efficient, and sustainable maritime transportation in the future.